\documentclass{mscs}
\usepackage{graphicx}
\usepackage{amsmath,amssymb} 
\newcommand{\ket}[1]{ \left |#1\right \rangle}
\newcommand{\bra}[1]{ \left \langle#1\right | }
\newcommand{\ip}[2]{\left\langle #1| #2 \right \rangle}
\newcommand{\op}[2]{\left | #1 \rangle\langle #2 \right |}
\newtheorem{lemma}{Lemma}
\newtheorem{corollary}{Corollary}
\newtheorem{property}{Property}
\newtheorem{theorem}{Theorem}
\newtheorem{definition}{Definition}
\newtheorem{example}{Example}

\begin{document}
  \title{Classically-controlled Quantum Computation} \author[Simon Perdrix and Philippe Jorrand]{Simon Perdrix and Philippe Jorrand\\IMAG, Universities of Grenoble, France\\
        Email: {simon.perdrix@imag.fr, philippe.jorrand@imag.fr}}
        \date{15 October 2005; Revised 27 March 2006}
        \maketitle

\begin{abstract} 

It is reasonable to assume that quantum computations take place under the control of the classical world. For modelling this standard situation, we introduce a Classically-controlled Quantum Turing Machine (CQTM) which is a Turing machine with a quantum tape for acting on quantum data, and a classical transition function for a formalized classical control. In a CQTM, unitary transformations and quantum measurements are allowed. We show that any classical Turing machine is simulated by a CQTM without loss of efficiency. Furthermore, we show that any $k$-tape CQTM is simulated by a $2$-tape CQTM with a quadratic loss of efficiency. In order to compare CQTMs to existing existing models of quantum computation, we prove that any uniform family of quantum circuits \cite{Y93} is efficiently approximated by a CQTM. Moreover we prove that any semi-uniform family of quantum circuits \cite{NO02}, and any measurement calculus pattern \cite{DK04} are efficiently simulated by a CQTM.  
Finally, we introduce a Measurement-based Quantum Turing Machine (MQTM) which is a restriction of CQTMs where only projective measurements are allowed. We prove that any CQTM is efficiently simulated by a MQTM. 
In order to appreciate the similarity between programming classical Turing machines and programming CQTMs, some examples of CQTMs are given.

\end{abstract}

\section{Introduction}\label{sec:intro}
Quantum computations operate in the quantum world. For their results to be useful in any way, by means of measurements for example, they operate under the control of the classical world. Quantum teleportation \cite{B93} illustrates the importance of classical control: the correcting Pauli operation applied at the end is classically controlled by the outcome of a previous measurement. 
Another example of the importance of classical control is measurement-based quantum computation \cite{L03,N01,P04,R03,DK04}, where classical conditional structures are required for controlling the computation. This classical control may be described as follows: \emph{``if the classical outcome of measurement number $i$ is $r_0$, then measurement number $i+1$ is on qubit $q_a$ according to observable $O_a$, otherwise measurement number $i+1$ is on qubit $q_b$ according to observable $O_b$"}. A particularly elegant formalization of measurement-based quantum computation is the measurement calculus \cite{DK04}.

The necessity of integrating the classical control in the description of quantum computations is a now well understood requirement in the design of high level languages for quantum programming \cite{JL03,S03}. There are also some proposals for lower level models of quantum computation integrating classical control, like the quantum random access machines (QRAM) \cite{K96,BSC01}. However there exist no formal and abstract model of quantum computation integrating classical control explicitly. This paper aims at defining such an abstract model of classically-controlled quantum computation.

One of the main existing abstract models of quantum computation is the Quantum Turing Machine (QTM) introduced by Deutsch \cite{D85}, which is an analogue of the classical Turing machine (TM). It has been extensively studied by Bernstein and Vazirani \cite{BV97}:  
a quantum Turing machine is an abstract model of quantum computers, which expands the classical model of a Turing machine by allowing a \emph{quantum} transition function. In a QTM, superpositions and interferences of configurations are allowed, but the classical control of computations is not formalized and inputs and outputs of the machine are still classical.
This last point means that the model of QTM explores the computational power of quantum mechanics for solving classical problems, without considering \emph{quantum} problems, i.e. quantum input/output.

While models dealing with quantum states like quantum circuits \cite{KSV,Y93} and QRAM, are mainly used for describing specific algorithms, the development of complexity classes, like $QMA$ \cite{W00}, which deal with quantum states, points out the necessity of theoretical models of quantum computation acting on quantum data.

The recently introduced Linear Quantum Turing Machine (LQTM) by S. Iriyama, M. Ohya, and I. Volovich \cite{IOV04} is a generalization of QTM dealing with mixed states and  irreversible transition functions which allow the representation of quantum measurements without classical outcomes. A consequence of this lack of classical outcome is that  the classical control is not formalized in LQTM and, among others, schemes like teleportation cannot be expressed. Moreover, similarly to  QTM, LQTM deals with classical input/output only.

We introduce here a Classically-controlled Quantum Turing Machine (CQTM) which is a TM with a quantum tape for acting on quantum data, and a classical transition function for a formalized classical control. In a CQTM, unitary transformations and quantum measurements are allowed. 
Theorem \ref{thm:tm} shows that any TM is simulated by a CQTM without loss of efficiency. 
In section \ref{sec:mtcqtm}, a CQTM with multiple tapes is introduced. Theorem \ref{thm:cqtm} shows that any $k$-tape CQTM is simulated by a $2$-tape CQTM with a quadratic loss of efficiency.
Moreover, a gap between classical and quantum computations is pointed out. 
In section \ref{sec:other}, the CQTM model is compared to two different models of quantum computation: the quantum circuit model \cite{Y93} and the measurement calculus \cite{DK04}. Both of them are efficiently simulated by CQTMs.
In section \ref{sec:mqtm}, a restriction of CQTMs to measurement-based quantum Turing machine is presented. In a MQTM only projective measurements are allowed. Theorem \ref{thm:cqtmmqtm} shows that any CQTM is simulated by a MQTM without loss of efficiency. 
To appreciate the similarity between programming a TM and programming a CQTM, some examples of CQTMs are given for solving problems like the recognition of quantum palindromes and the insertion of a blank symbol in the input data.
 A perspective is to make the CQTM not only a well defined theoretical model but also a bridge to practical models of quantum computations like QRAM, by relying on the fact that natural models of quantum computations are classically controlled.

\section{Quantum Computing Basics}
\label{sec:basics}

\subsection{Quantum states}

The basic carrier of information in quantum computing is a $2$-level quantum system (\emph{qubit}), or more generally a $d$-level quantum system (\emph{qudit}). The state of a single qudit is a normalized vector of the $d$-dimensional Hilbert space $\mathbb C^d$. An orthonormal  basis (o.n.b.) of this Hilbert space is described as $\{\ket{\tau}, \tau \in \Sigma_Q\}$, where $\Sigma_Q$ is a finite alphabet of symbols such that $\vert \Sigma_Q\vert  = d$. So the general state $\ket \phi \in \mathcal H_{\Sigma_Q}$ of a single qudit can be written as $$\sum_{\tau \in \Sigma_Q} \alpha_{\tau} \ket{\tau},$$ with $\sum_{\tau \in \Sigma_Q} \vert \alpha_{\tau} \vert ^2=1$.

Vectors, inner and outer products are expressed in the notation introduced by Dirac. Vectors are denoted $\ket \phi$ ; the inner product of two vectors $\ket \phi$, $\ket \psi$ is denoted by $\ip{\phi}{\psi}$. If $\ket{\phi} = \sum_{\tau\in \Sigma_Q} \alpha_\tau \ket \tau$ and $\ket{\psi} = \sum_{\tau\in \Sigma_Q} \beta_\tau \ket \tau$, then $\ip \phi\psi= \sum_{\tau \in \Sigma_Q} \alpha_\tau^*\beta_\tau$ (where $\alpha^*$ stands for the complex conjugate). 

The left hand side $\bra \phi$ of the inner product is a \emph{bra-vector}, and the right hand side $\ket \psi$ is a \emph{ket-vector}. A bra-vector is defined as the adjoint of the corresponding ket-vector:
If $\ket \phi = \sum_{\tau\in \Sigma_Q} \alpha_\tau \ket \tau$, then $\bra \phi  =\ket \phi ^\dagger = \sum_{\tau\in \Sigma_Q} \alpha_\tau^* \bra \tau$.

The bra-ket  notation can also be used to describe outer products:  $\op \phi \psi$ is a linear operator, $(\op \phi \psi)\ket \epsilon = \ip \psi \epsilon \ket \phi$.

The state of a system of $n$ qudits is a normalized vector in $\otimes_{i=1}^n\mathbb C^d \cong \mathbb C^{d^n}$, where $\otimes$ is the tensor product of vector spaces. $\ket {\tau \gamma}$ denotes $\ket \tau \otimes \ket \gamma$, such that a basis vector of $\mathbb C^{d^n}$ can be denoted $\ket \omega$, where $\omega \in \Sigma_Q^n$. As a special case, if $n=0$, the basis vector in $\mathbb C^1$ is denoted $\ket {}$ \cite{SV05}. Notice that for any $n>0$, $\mathbb C^n\otimes\mathbb C^1 \cong \mathbb C^n$.

\subsection{Quantum evolutions}

The three basic operations in quantum computing are \emph{unitary transformation}, \emph{initialization}, and \emph{measurement}.
\begin{itemize}
\item A unitary operation maps an $n$-qudit state to a $n$-qudit state, and is given by a unitary $d^n\times d^n$-matrix $U$. This unitary operation transforms $\ket {\phi}$ into $U\ket{\phi}$. 
\item Initializing a qudit according to a special state, say $\ket {\tau_0}$, maps a $0$-qudit state to a $1$-qudit state, and is given by the matrix $\op{\tau_0}{}$. 
 This initialization transforms $\ket{}$ into $\ket{\tau_0}\ip{} {} = \ket{\tau_0}$.

\item Two kinds of measurements are considered: 
\subitem A destructive measurement according to an o.n.b. $\{\ket{\tau}, \tau \in \Sigma_Q\}$, maps a $1$-qudit state to a $0$-qudit state. If the state is $\ket{\phi}$ immediatly before the measurement then the probability that the classical result $\tau\in \Sigma_Q$ occurs is $p(\tau)= \vert \ip \tau \phi\vert ^2$, and the state of the system after the measurement is $\ket {}$. 
\subitem A projective measurement maps a $n$-qudit state to a $n$-qudit state, and is given by a collection of $d^n\times d^n$-matrices $\{P_k\}_{ k\in I}$, such that $P_kP_l=\delta_{k,l}P_k$ and $\sum_{k\in I} P_k= Id$. Any projective measurement $\{P_k\}_{k\in I}$ can be characterized by an \emph{observable} $O=\sum_{k\in I} \alpha_k P_k$, for some distinct $\alpha_k \in \mathbb R$.
If the state is $\ket{\phi}$ immediatly before the measurement then the probability that the classical result $k \in I$ occurs is $p(k)=  \bra \phi P_k\ket \phi$, and the state of the system after the measurement is $\frac{P_k\ket{\phi}}{\sqrt{p(k)}}$. 

\end{itemize}

Unitary operations can be spatially composed by means of tensor product: if $U$  is a $n$-qudit unitary operation  and $V$ is a $m$-qudit operation, then $U\otimes V$ is a $n+m$-qudit unitary transformation.
Unitary operation can also be sequentially composed by means of matrix product: if $U$ and $V$ are two $n$-qudit operations, then $V.U$ is a $n$-qudit unitary transformation consisting in applying $U$ and then $V$.

The traditional scheme of quantum computation consists in initializing some qudits, then in applying unitary operations, and finally in performing a destructive measurement of each qudit of the system. In this traditional scheme, computations can be described by means of the quantum circuit model \cite{Y93}.

Recent alternative models of quantum computation \cite{DK04,P04,R03}, do not follow this traditional scheme, allowing for instance sequential composition of projective measurements. Since projective measurements are not closed under sequential composition, a more general formalism, called \emph{admissible transformations} or \emph{general measurements} is used to describe all the basic quantum operations (unitary operation, initialization, measurements). Moreover this formalism  is closed under spatial and sequential compositions.

\begin{definition}[Admissible transformation] \label{def:at}
An admissible transformation of a $n$-qudit state into a $m$-qudit state is described by a collection $\{M_\tau, \tau \in \Sigma_C\}$ of linear operators mapping $\mathbb C^{d^n}$ to $\mathbb C^{d^m}$  satisfying the completness equation $$\sum_{\tau \in \Sigma_C} M_\tau^\dagger M_\tau = Id_{\mathbb C^{d^n}}$$ where $\Sigma_C$ is a finite set of classical outcomes.

 If the state of the quantum system is $\ket{\psi}$ immediately before the transformation then the probability that the classical outcome $\tau\in \Sigma_C$ occurs is given by $$p(\tau)= \langle \psi \vert M_{\tau}^{\dagger} M_{\tau}^{ }\ket{\psi},$$ and the state of the system after the transformation is $$\frac{M_{\tau}\ket{\psi}}{\sqrt{p(\tau)}}.$$
\end{definition}

\begin{property}[Sequential Composition]
Let $T$ be an admissible transformation of a $n$-qudit state into a $m$-qudit state, described by $\{M_\tau, \tau \in \Sigma_C\}$, and $T'$  be an admissible transformation transforming a $m$-qudit state into a $k$-qudit state, described by $\{N_\gamma, \gamma \in \Sigma'_C\}$.
 The sequential composition of $T$ and $T'$ is an admissible transformation $\tilde T$ transforming a $n$-qudit state into a $k$-qudit state, described by $\{N_\gamma M_{\tau}, (\tau,\gamma) \in \Sigma_C\times \Sigma'_C\}$. \end{property}

\begin{property}[Spatial Composition]
Let $T$ be an admissible transformation of a $n$-qudit state into a $m$-qudit state, described by $\{M_\tau, \tau \in \Sigma_C\}$, and $T'$  be an admissible transformation of a $n'$-qudit state into a $m'$-qudit state, described by $\{N_\gamma, \gamma \in \Sigma'_C\}$.
 The spatial composition of $T$ and $T'$ is an admissible transformation $\tilde T$ of a $(n+n')$-qudit state into a $(m+m')$-qudit state, described by $\{M_{\tau}\otimes N_\gamma, (\tau,\gamma) \in \Sigma_C\times \Sigma'_C\}$.
\end{property}

All basic quantum operations can be described by means of admissible transformations: 
\begin{itemize}
\item A unitary operation $U$ is nothing but an admissible transformation $\{M_\lambda\}$ where $M_\lambda =U$. The completeness equation is satisfied and the classical outcome $\lambda$ occurs with probability $1$, where $\lambda$ is the void classical outcome.
\item A qudit initialization according to $\ket{\tau_0}$ is described by $\{ M_\lambda \}$ where $M_\lambda = \ket{\tau_0}\bra {}$. Since $(\ket{\tau_0} \bra{})^\dagger \ket{\tau_0}\bra{} = \ket{} \ip{\tau_0}{\tau_0}\bra {}=\ket{}\bra{} = Id_{\mathbb C^1}$, the completeness equation is satisfied. Moreover the classical outcome $\lambda$ occurs with probability $1$.
\item A destructive measurement according to an o.n.b. $\{\ket{\tau}, \tau \in \Sigma_Q\}$ is described by $\{M_\tau,\tau \in \Sigma_Q \}$ where $M_\tau = \ket {}\bra{\tau}$. Since $\{\ket{\tau}, \tau \in \Sigma_Q\}$ is an o.n.b,  $\sum_{\tau \in \Sigma_Q} \ket{\tau}\ip {}{}\bra{\tau}=\sum_{\tau \in \Sigma_Q} \ket{\tau}\bra{\tau}=Id_{\mathbb C^{\vert \Sigma_Q\vert}}$: the completeness equation is satisfied. Moreover if the state of the qudit is $\ket \psi$ just before the measurement, the probability that the classical outcome $\tau \in \Sigma_Q$ occurs is $p(\tau)=\vert \ip \tau \psi\vert ^2=\ip \psi \tau\ip\tau\psi=\bra \psi\ket\tau\bra {}\ket {}\bra\tau \ket\psi=\bra \psi M_\lambda^\dagger M_\lambda \ket \psi$. The state of the system after the measurement is $\frac{M_{\tau}\ket{\psi}}{\sqrt{p(\tau)}}= \ket {}$
\item A projective measurement described by $\{P_k, k\in I\}$ is an admissible transformation described by the same collection of linear operators. 
\end{itemize}

Admissible transformations allow the  representation of the basic quantum operations and are closed under sequential and spatial compositions. But one can wonder whether all the admissible transformations have a physical meaning. It turns out that any admissible transformation can be simulated in the traditional scheme of quantum computation consisting in ($i$) initialization, ($ii$) sequence of unitary transformations, and ($iii$) destructive measurement \cite{NC00}.

One can imagine a generalized quantum circuit model where unitary transformations are replaced by admissible transformations. But, contrary to unitary transformations, admissible transformations produce a classical result which allows a classical control consisting, for instance, in conditional compositions 
 and loops. The classically controlled quantum Turing machine  is a new model of quantum computation which takes classical control into account.
 
In the following, some basic admissible transformations will be largely used: 
For a given Hilbert space $\mathcal{H}_{\Sigma_Q}$, we exhibit some admissible transformations with classical results belonging to a finite set $\Sigma_C=\Sigma_Q \cup \overline{\Sigma_Q} \cup \{\lambda, \top, \bot\}$, where $\overline{\Sigma_Q}=\{\overline{\tau}:\tau \in \Sigma_Q\}$ and $\lambda, \top, \bot \notin \Sigma_Q$:

\begin{itemize}
\item $Std=\{M_{\tau}\}_{\tau \in \Sigma_Q} $ is a projective measurement in the standard basis: $\forall \tau \in \Sigma_Q, M_{\tau}=\ket{\tau}\langle \tauÊ\vert$,
\item $\mathcal{T}_{\tau}=\{M_{\tau},M_{\overline{\tau}}\}$ is a test for the symbol $\tau$: $M_{\tau}=\ket{\tau}\langle \tauÊ\vert$ and $M_{\overline{\tau}}=I-\ket{{\tau}}\langle {\tau}Ê\vert$,
\item $\mathcal{P}_{[\tau_a,\tau_b]}=\{M_{\lambda}\}$ is a unitary transformation with outcome $\lambda$, and $M_{\lambda}= \ket{{\tau_a}}\langle {\tau_b}Ê\vert+\ket{{\tau_b}}\langle {\tau_a}Ê\vert + (\sum_{\tau \in \Sigma_Q - \{\tau_a,\tau_b\}} \ket{{\tau}}\langle {\tau}Ê\vert)$ is a permutation of the symbols $\tau_a$ and $\tau_b$.
\item $Swap=\{M_\lambda\}$ is a $2$-qudit unitary operation with outcome $\lambda$, swapping the state of the qudits: $M_\lambda = \sum_{\tau,\gamma\in \Sigma_Q}\ket{\gamma\tau}\bra{\tau \gamma}$.

\item $\mathcal{U}_V=\{M_{\lambda}\}$ is the unitary transformation $M_{\lambda}=V$, with classical outcome $\lambda$.

\item $\mathcal{O}_O=\{P_{k}\}_k$, is a projective measurement according to the observable $O=\Sigma_k \alpha_k P_k$. 
\item $\mathcal{O}_{[\tau_a,\tau_b]}=\{P_\top, P_\bot\} \cup \{P_\tau\}_{\tau \in \Sigma_C -\{\tau_a, \tau _b\}}$ is a projective measurement in a basis \emph{diagonal} according to $\tau_a$, $\tau_b$ : $\forall {\tau \in \Sigma_C -\{\tau_a, \tau _b\}}$, $P_\tau = \ket \tau \bra\tau$, $P_\top = (\ket{\tau_a} + \ket{\tau_b}) (\bra{\tau_a} + \bra{\tau_b})/2$, and $P_\bot = (\ket{\tau_a} - \ket{\tau_b}) (\bra{\tau_a} - \bra{\tau_b})/2$.

\end{itemize}

  \section{Classically-controlled Quantum Turing Machines}\label{sec:cqtm}
 
 For completeness, definition \ref{def:tm} is  the definition of a deterministic TM \cite{P94}. A classically-controlled quantum Turing machine (definition \ref{def:cqtm}) is composed of a quantum tape of quantum cells, a set of classical internal states and a head for applying admissible transformations to cells on the tape. The role of the head is crucial because it implements the interaction across the boundary between the quantum and the classical parts of the machine.

 \begin{definition}  \label{def:tm} A deterministic (classical) Turing Machine is defined by a triplet $M=(K,\Sigma, \delta)$, where $K$ is a finite set of states with an identified initial state $s$, $\Sigma$ is a finite alphabet with an identified ``blank" symbol $\#$, and $\delta$ is a deterministic transition: $$\delta: K\times \Sigma \to (K\cup \{ ``yes", ``no", h\})\times \Sigma \times \{ \leftarrow, \rightarrow, - \}. $$ We assume that $h$ (the halting state), $``yes"$ (the accepting state) and $``no"$ (the rejecting state) are not in $K$.
 \end{definition}

 \begin{definition} \label{def:cqtm} A Classically-controlled Quantum Turing Machine is a quintuple $M=( K,
\Sigma_{{C}}, \Sigma_{Q},\mathcal{A},\delta)$. Here $K$ is a finite set of classical states with an identified initial state $s$, 
$\Sigma_{{Q}}$ is a finite alphabet which denotes basis states of quantum cells, 
$\Sigma_{{C}}$ is a finite alphabet of classical outcomes, $\mathcal{A}$ is a finite set of one-quantum cell admissible transformations,
and $\delta$ is a classical transition function: 
$$\delta:K\times\Sigma_{C} \to  (K\cup \{ ``yes", ``no", h\})  \times \{ \leftarrow, \rightarrow, - \}\times \mathcal{A}.$$ We assume that $h$ (the halting state), $``yes"$ (the accepting state) and $``no"$ (the rejecting state) are not in $K$, and that all possible classical outcomes of each transformation of $\mathcal{A}$ are in $\Sigma_{C}$. Moreover we assume that 
$\Sigma_Q$ always contains a ``blank" symbol $\#$, $\Sigma_C$ always contains a ``blank" symbol $\#$ and a ``non-blank" symbol $\overline{\#}$, and $\mathcal{A}$ always contains the admissible ``blank test" transformation $\mathcal{T}_{\#}$.
 \end{definition} 

\par{The function $\delta$ is a formalization of the classical control of  quantum computations and can also be viewed as the ``program" of the machine. It specifies, for each combination of current classical state $q\in K$ and last obtained classical outcome $\tau \in \Sigma_C$, a triplet $\delta(q, \tau)=(p, D,A)$, where $p$ is the next classical state, $D \in \{ \leftarrow, \rightarrow, - \}$ is the direction in which the head will move, and $A\in \mathcal{A}$ is the admissible transformation to be performed next. The \emph{blank test} admissible transformation $\{M_{\#},M_{\overline{\#}}\}$ establishes a correspondence between the quantum blank symbol ($\#$) and the classical blank ($\#$) and non-blank ($\overline{\#}$) symbols: if the state $\ket{\phi}$ of the measured quantum cell is $\ket{\#}$, the outcome of the measurement is $\#$ whereas if $\ket{\phi}$ is orthogonal to $\ket{\#}$ ($\langle \phi\ket{\#}=0$) then the outcome is $\overline{\#}$.}
 
\par{ How does the program start? The generally unknown quantum input of the computation $\ket{\phi}= $ $\sum_{\tau \in (\Sigma_Q-\{\#\})^n} \alpha_{\tau} \ket{\tau}$  
 is placed on $n$ adjacent cells of the tape, while the state of all other quantum cells of the tape is $\ket{\#}$. The head is pointing at the blank cell immediately located on the left of the input. Initially, the classical state of the machine is $s$ and $\#$ is considered as the last classical outcome, thus the first transition is always $\delta(s,\#)$.}

\par{ How does the program halt? The transition function $\delta$ is total on $K\times \Sigma_C$ (irrelevant transitions will be omitted from its description). There is only one reason why the machine cannot continue: one of the three halting states $h$, $``yes"$, and $``no"$ has been reached. 
 If a machine $M$ halts on input $\ket{\phi_{in}}$, the output $M(\ket{\phi_{in}})$ of the machine $M$ on $\ket{\phi_{in}}$ is defined. If states $``yes"$ or $``no"$ are reached, then $M(\ket{\phi_{in}})=``yes"$ or $``no"$ respectively. Otherwise, if halting state $h$ is reached then the output is the state $\ket{\phi_{out}}$ of the tape of $M$ at the time of halting. Since the computation has gone on for finitely many steps, only a finite number of cells are not in the state $\ket{\#}$. The output state $\ket{\phi_{out}}$ is the state of the finite register composed of the quantum cells from the leftmost cell in a state which is not $\ket{\#}$ to the rightmost cell in a state which is not $\ket{\#}$. Naturally, it is possible that $M$ never halts on input $\ket{\phi_{in}}$. If this is the case we write $M(\ket{\phi_{in}})=\nearrow$.}

 Since quantum measurement is probabilistic, for a given input state $\ket{\phi_{in}}$, a CQTM does not, in general, always produce the same output, so there exists a probability distribution over possible outputs. Moreover the halting time of a CQTM $M$ on an input $\ket{\phi_{in}}$
  is also a probability distribution. Thus two special classes of CQTMs can be distinguished: \emph{Monte Carlo} and \emph{Las Vegas}. For a given CQTM $M$, if for a given input $\ket{\phi_{in}}$
   there exists a finite and non-probabilistic bound for the execution time of $M$, then $M$ is \emph{Monte Carlo}. If the output $M(\ket{\phi_{in}})$ is not probabilistic 
 then $M$ is \emph{Las Vegas}. An example of \emph{Monte Carlo} CQTM is given in example \ref{ex:pal}: this CQTM recognizes a language composed of "quantum palindromes", i.e. quantum states which are superpositions of palindromes. An example of \emph{Las Vegas} CQTM which simulates the application of a given 1-qubit unitary transformation $U$ ($H$ in the example) on a quantum state using projective measurements only is given in example \ref{ex:h}. In section \ref{sec:tm}, we use a CQTM which is both \emph{Las Vegas} and \emph{Monte Carlo} for simulating a classical TM.

A \emph{configuration} of a CQTM $M$ is a complete description of the current state of the computation. Formally, a configuration is a triplet $(q,\tau, \ket{\psi})$, where $q\in K\cup\{h,``yes",``no"\}$ is the internal state of $M$, $\tau \in \Sigma_C$ is the last obtained outcome, and $\ket{\psi} \in \mathcal{H}_{\Sigma_Q'}$ represents the state of the tape and the position of the head. Here $\Sigma_Q'=\Sigma_Q\cup \underline{\Sigma}_Q$, where $\underline{\Sigma}_Q=\{\underline{\tau} : \tau \in \Sigma_Q\}$ is a set of \emph{pointed} versions of the symbols in $\Sigma_Q$: 
if $\ket \phi \in \mathcal H_{\Sigma_Q}$ is the state of the tape, and if the head is pointing at cell number $k$, $\ket{\psi}\in \mathcal H_{\Sigma_Q'}$ is obtained by replacing all symbols $x\in \Sigma_Q$ at the $k^\text{th}$ position in $\ket \phi$ by the corresponding $\underline x \in \Sigma_Q'$.
 For instance, if $K=\{q_1,q_2\}$, $\Sigma_C=\{\#, \overline{\#}, t,u,v\}$ and $\Sigma_Q=\{\#, a,b\}$, the configuration $$(q_1,u,\frac{1}{\sqrt{2}} (\ket{a\#\underline{b}b}+\ket{b\#\underline{a}b}))$$ means that the internal state of the machine is $q_1$, the last outcome is $u$, the state of the tape is $\frac{1}{\sqrt{2}}(\ket{a\# bb}+\ket{b\# ab})$, and the head is pointing at the third cell from the left.

\begin{example}[Quantum palindromes]\label{ex:pal} Consider the CQTM $M=(K, \Sigma_C, \Sigma_Q, \mathcal{A}, \delta)$, 
with $K=\{s, q, q_0, q_1,$ $ q_0', q_1', \tilde{q}\}$, $\Sigma_C=\{\#,\overline{\#}, 0,1,\lambda\}$, 
$\Sigma_Q=\{\#,0,1\}$ and 
$\mathcal{A}=\{\mathcal{T}_{\#},Std,$ $\mathcal{P}_{[0,\#]},\mathcal{P}_{[1,\#]}\}$ (these admissible transformations are defined in section \ref{sec:basics}), and $\delta$ as described in figure \ref{fig:pal}.

The purpose of this machine is to tell whether its input is a quantum palindrome, i.e. a state which is a superposition of basis states such that each basis state in the superposition is a palindrome. For instance the states: 
$\ket{00}$,
 $\frac{1}{\sqrt{2}}(\ket{010}+i\ket{111})$,
 $\frac{1}{\sqrt{2}}(\ket{00000}+\ket{11111})$ are quantum palindromes. The machine works as follows: the first cell of the input is measured in the standard basis and replaced with $\ket{\#}$, the result is memorized by means of the internal states $q_0$ and $q_1$, then $M$ moves right, up to the end of the input. The last cell is then measured in the standard basis: if the outcome agrees with the one remembered, it is replaced with $\ket{\#}$.  $M$ then moves back left to the beginning of the remaining input and the process is repeated. The transition function is described in figure  \ref{fig:pal}. For instance, if the internal state is $q_0$ and the last obtained classical outcome is $\#$, then the internal state becomes $q'_0$, the head moves to the left and then the pointed cell is measured in the standard basis.
 
 This machine is a Monte Carlo CQTM operating in time $O(n^2)$, where $n$ is the size of the input. Considering the language $L\subset \mathcal{H}_{\Sigma_Q}$ composed of quantum palindrome states, if $\ket{\phi_{in}}\in L$, then the probability that $M$ accepts $\ket{\phi_{in}}$ is $Pr[M(\ket{\phi_{in}})=``yes"]=1$: if the input is a quantum palindrome then, in any case, the machine recognizes $\ket{\phi_{in}}$, but $M$ may accept states which are not palindromes with high probability, for instance
  $\forall \epsilon > 0, Pr[M(\sqrt{1-\epsilon}\ket{00}+\sqrt{\epsilon}\ket{10})=``yes"]=1-\epsilon$.

 \begin{figure*}
 \begin{center}
 \begin{tabular}{ccc}
 $\begin{array}{ccc}
 \hline
 p\in K,&\ \tau \in \Sigma_C&\delta(p,\tau)\\
 \hline
 s & \# & (q,\rightarrow,Std ) \\
  q & \# & (``yes",-,- ) \\
 q & 0 & (q_0,-,\mathcal{P}_{[0,\#]} ) \\
 q & 1 & (q_1,-,\mathcal{P}_{[1,\#]}) \\
 q_0 & \lambda & (q_0,\rightarrow,\mathcal{T}_{\#} ) \\
q_0 & \overline{\#} & (q_0,\rightarrow,\mathcal{T}_{\#}) \\
q_0 & \# & (q_0',\leftarrow,Std ) \\
   q_1 & \lambda & (q_1,\rightarrow,\mathcal{T}_{\#} ) \\
   q_1 & \overline{\#} & (q_1,\rightarrow,\mathcal{T}_{\#} ) \\
  q_1 & \# & (q_1',\leftarrow,Std ) \\
 \hline
 \end{array}$&$\ \ $
 $\begin{array}{ccc}
 \hline
 p\in K&\tau \in \Sigma_C&\delta(p,\tau)\\
 \hline
    q_0' & \# & (``yes",-,- ) \\
 q_0' & 0 & (\tilde{q},-,\mathcal{P}_{[0,\#]} ) \\
  q_0' & 1 & ("no",-,- ) \\ 
  q_1' & \# & (``yes",-,- ) \\
 q_1' & 0 & ("no",-,- ) \\
  q_1' & 1 & (\tilde{q},-,\mathcal{P}_{[1,\#]}) \\
\tilde{q} & \lambda & (\tilde{q},\leftarrow,\mathcal{T}_{\#} ) \\
   \tilde{q} & \overline{\#} & (\tilde{q},\leftarrow ,\mathcal{T}_{\#}) \\
\tilde{q} & \# & (q,\rightarrow,Std ) \\
 \hline
 \end{array}$
 \end{tabular}
  
 \caption{CQTM for quantum palindromes. The symbol $"-"$ used as an admissible transformation, means $\mathcal{U}_I$, i.e. the identity transformation with $\lambda$ as classical outcome.}\label{fig:pal}
 \end{center}
 
 \end{figure*}
 \end{example}

 \section{CQTM and TM}
 \label{sec:tm}
 
 The following theorem shows that any TM is simulated by a CQTM without loss of efficiency. 
 
 \begin{theorem}\label{thm:tm} Given any TM $M_C$ operating in time $f(n)$, where $n$ is the input size, there exists a CQTM $M_Q$ operating in time $O(f(n))$ and such that for any input $x$, $M_C(x)=M_Q(\ket{x})$. \footnote{If the halting state $h$ is reached, $M_Q(\ket x)$ denotes the final state of the tape. So, if $h$ is reached, $M_C(x)=M_Q(\ket x)$ has to be replaced by $M_Q(\ket x)=\ket{M_C(x)}$.} 
 \end{theorem}
 
  \begin{proof}
 For a given TM $M_C=(K,\Sigma,\delta_C)$, we describe a CQTM $M_Q$ which simulates $M_C$. One way to do this is to simulate the classical tape of $M_C$ using only basis states of the quantum tape of $M_Q$. 

Formally, we consider the CQTM $M_Q=(K\cup K_{\Sigma}\cup\{s'\},\Sigma\cup \{\overline{\#},\lambda\},\Sigma, \mathcal{A},\delta_Q)$.
 Here $K_{\Sigma}=\{q_{\tau}:q\in K, \tau \in \Sigma\}$, $\mathcal{A}=\{Std\}\cup \{\mathcal{P}_{[\tau_1,\tau_2]}\}_{\tau_1,\tau_2 \in \Sigma}$. The initial state of $M_Q$ is $s'$ and its first transition is $\delta_Q(s',\#)=(s,-,Std)$, where $s$ is the initial state of $M_C$.
 For any $(q,\tau) \in K\times \Sigma$, the transition $\delta_C(q,\tau)=(q',\tau',D)$ is decomposed into two transitions:
  $\delta_Q(q,\tau)=(q_{\tau},-,\mathcal{P}_{[\tau,\tau']})$ and 
  $\delta_Q(q_{\tau},\lambda)=(q',D,Std)$.

Since each transition of $M_C$ is simulated with probability $1$ by two transitions of $M_Q$, if $M_C$ operates in time $f(n)$, $M_Q$ operates in time $2f(n)$, where $n$ is the size of the input. \end{proof}
 
 Any TM is simulated by a CQTM without loss of efficiency. 
 However, as will be shown in lemma \ref{lem:nuniv}, a CQTM with one tape cannot simulate some other models of quantum computation, like quantum circuits, because only one-cell admissible transformations are allowed. In order to allow transformations on more than one cell, we introduce multi-tape CQTMs. With $k$ heads, $k$-cell admissible transformations can be performed.

 \section{CQTM with multiple tapes}
 \label{sec:mtcqtm}

 We show that any $k$-tape CQTM is simulated by a $2$-tape CQTM with an inconsequential loss of efficiency. Moreover, by showing that $1$- and $2$-tape CQTM are not equivalent, we point out a \emph{gap} between classical and quantum computations.
 
 \begin{definition} \label{def:ktape} A $k$-tape classically-controlled quantum Turing machine where $k>0$, is a quintuple $M=( K ,\Sigma_{{C}}, \Sigma_{Q},\mathcal{A},\delta)$, where $K$ is a finite set of classical states with an identified initial state $s$, $\Sigma_Q$ is a finite alphabet which denotes basis states of each quantum cell, $\mathcal{A}$ is a finite set of $k$-cell admissible transformations, $\Sigma_C$ is a finite alphabet of classical outcomes of $k$-cell admissible transformations and $\delta$ is a classical transition function 
$$\delta:K\times\Sigma_{C} \to  (K\cup \{ ``yes", ``no", h\})\times (\{ \leftarrow, \rightarrow, - \})^k  \times \mathcal{A}.$$
We assume that all possible classical outcomes of each measurement of $\mathcal{A}$ are in $\Sigma_{C}$ and that $\mathcal{A}$ always contains the $k$ admissible ``blank test" transformations, one for each tape of the machine.
 \end{definition} 

 Intuitively, $\delta(q,\tau)=(q',(D_1,\ldots D_k),A)$ means that, if $M$ is in state $q$ and the last classical outcome is $\tau$, then the next state will be $q'$, the $k$ heads of the machine will move according to $D_1,\ldots ,D_k$ and the next $k$-quantum cell admissible transformation will be $A$. This admissible transformation will be performed on the $k$ quantum cells pointed at by the heads of the machine after they have moved. A $k$-cell admissible transformation $A$ can be defined directly, for instance by use of a $k$-cell unitary transformation $V$ ($A=\mathcal{U}_V$). $A$ can also be defined as a composition of two admissible transformations $A_1$, $A_2$ respectively on $j$ and $l$ cells such that $j+l=k$, then $A=[A_1,A_2]$ means that the first $j$ heads apply $A_1$ and, simultaneously, the last $l$ heads apply $A_2$. The classical outcome is the concatenation of the outcomes of $A_1$ and $A_2$, where $\lambda$ is the unit element of the concatenation (i.e. $\tau.\lambda=\tau$).

 A $k$-tape CQTM starts with an input state $\ket{\phi}$ on a specified tape $T_1$, with all cells of other tapes in state $\ket \#$, and if the halting state $h$ is reached, the machine halts and the output is the state of the specified tape $T_1$.

  \begin{example}[Inserting a blank symbol]\label{ex:blank}
 Consider the problem of inserting a blank symbol between the first and the second cells of a quantum state $\ket{\psi_{in}}$ which resides on one of the tapes. For instance, $\ket{abba}$ is transformed into $\ket{a\#bba}$, and $\frac{1}{\sqrt{2}}(\ket{aa}+\ket{bb})$ into $\frac{1}{\sqrt{2}}(\ket{a\#a}+\ket{b\#b})$.
Consider the $2$-tape CQTM $M=(K, \Sigma_C, \Sigma_Q, \mathcal{A}, \delta)$, 
with $K=\{s, q_0, q_1\}$, 
$\Sigma_C=\{\#,\overline{\#}, \lambda \}$ and 
$\mathcal{A}=\{\mathcal T_\#, Swap\}$. 
$\delta$ is described in figure \ref{fig:blank}.
\begin{figure}
 \begin{center}
 $\begin{array}{ccc}
 \hline
 p\in K,&\ \tau \in \Sigma_C&\delta(p,\tau)\\
 \hline
 s & \# & (q_0,(\leftarrow,-), Swap) \\
  q_0 & \lambda & (q_1,(\rightarrow,-),Swap) \\
 q_1 & \lambda & (h,(\rightarrow,-),- ) \\
\hline
 \end{array}$
   \caption{$2$-tape CQTM for inserting a blank symbol}\label{fig:blank}
 \end{center}
  \end{figure}

The input state is on the first tape and let $a$ be a name for the first cell on the left of the input.
In order to insert a blank symbol in the second position of the input state, the state of $a $ 
is swapped with a cell of the second tape. Then the state of this cell on the second tape is swapped with the state of the cell immediatly located to the left of $a$. 
 \end{example}

\begin{theorem}\label{thm:cqtm}
Given any $k$-tape CQTM $M$ operating in time $f(n)$, where $n$ is the input size,
 there exists a $2$-tape CQTM $M'$ 
 operating in time $O(f(n)^2)$
  and such that for any input $\ket{\psi}, M(\ket{\psi})=M'(\ket{\psi})$.
\end{theorem}

\begin{proof} Suppose that $M=(K,\Sigma_C,\Sigma_Q, \mathcal{A},\delta)$ has $k$ tapes, we describe $M'=(K',\Sigma_C',$ $\Sigma_Q',\mathcal{A}',\delta')$ having only two tapes. $M'$ must "simulate" the $k$ tapes of $M$. One way to do this is to maintain on one tape $T_1$ of $M'$ the \emph{concatenation} of the contents of the tapes of $M$. The position of each head must also be remembered. 

To accomplish that, $\Sigma_Q'=\Sigma_Q \cup \underline{\Sigma}_Q \cup \{\triangleright, \triangleleft \}$, where $\underline{\Sigma}_Q=\{\underline{\tau}:\tau\in \Sigma_Q\}$ is a set of pointed versions of the symbols in $\Sigma_Q$, and $\triangleright$ ($\triangleleft$) signals the left (right) end of each simulated tape. Intuitively, at each step of the computation, if $\ket{\phi_j}$ is the state of each tape $j$ of $M$, the state of the tape $T_1$ of $M'$ is $\ket{\triangleright\triangleright}\ket{\phi_1}\ket{\triangleleft\triangleright}\ket{\phi_2}\ket{\triangleleft\triangleright}\ldots\ket{\triangleleft\triangleright}\ket{\phi_k}\ket{\triangleleft\triangleleft}$. In order to remember the positions of the $k$ heads, a unitary transformation is applied to the cells of $M'$ corresponding to cells of $M$ pointed at by the heads of $M$. This unitary transformation replaces
the symbols of $\Sigma_Q$ by their corresponding versions in $\underline{\Sigma}_Q$

Since each $k$-cell admissible transformation from $\mathcal{A}$ can be decomposed into $l_{\mathcal{A}}$ $2$-cell admissible transformations (see \cite{MS00}), $\mathcal{A'}$, which is composed of $1$- and $2$-cell admissible transformations, is defined such that any transformation from $\mathcal{A}$
 can be simulated with a finite number $l_{\mathcal{A}}$ 
 of transformations of $\mathcal{A'}$. 
 
 For the simulation to begin, $M'$ inserts a $\triangleright\triangleright$ to the left and 
$\triangleleft(\triangleright\triangleleft)^k\triangleleft$ to the right of the input, since the input of $M$ is located on its first tape. 
For simulating a transition $\delta(q,\tau)=(q',D,A)$ of $M$, the pointed cells change first according to $D$. Notice that if a head meets the symbol $\triangleright$,
  then a blank symbol is inserted to the right of this cell (see example \ref{ex:blank}) for simulating the infinity of the tapes, and similarly for the symbol $\triangleleft$.
 $A$ is simulated via a sequence of $2$-cell transformations. Since $2$-cell transformations are possible only on cells located on different tapes, the state of one of the two cells is transferred (by means of \emph{Swap}, see  example \ref{ex:blank}) from tape $T_1$ to the other tape $T_2$. Then the $2$-cell transformation is performed, and the state located on $T_2$ is transfered back to 
$T_1$, and so on. 
In order to reconstruct the classical outcome of the simulated transformation $A$, $M'$ must go through new internal states which keep track of the classical outcomes of the different $1$- and $2$-cell transformations.

 The simulation proceeds until $M$ halts. How long does the computation from an input $\ket{\phi}$ of size $n$ take? Since $M$ halts in time $f(n)$, no more than $k.f(n)$ cells of $M$ are non-blank cells. Thus the total length of the non-blank cells of $M'$ is $k.(f(n)+2)+3$ (to account for the $\triangleleft, \triangleright$ and the cell of $T_2$ used for the application of $2$-quantum cell transformations). Simulating a move of the heads takes at most two traversals of the non-blank cells of $T_1$. Each simulation of an admissible transformation of $\mathcal{A}$ 
requires a constant number $l_{\mathcal{A}}$ of transformations of $\mathcal{A'}$ ($l_{\mathcal{A}}$ is independent of the input size), 
moreover the simulation of each transformation in $\mathcal{A'}$ requires two traversals. As a consequence, the simulation of each transition of $M$ requires $O(f(n))$ transitions of $M'$, thus the total execution time of $M'$ is $O(f(n)^2)$. \end{proof}

The following lemma shows that some $2$-tape CQTMs cannot be simulated by $1$-tape CQTMs:

\begin{lemma}\label{lem:nuniv}
There exists a $2$-tape CQTM $M$ such that no $1$-tape CQTM simulates $M$.
\end{lemma}

\begin{proof}
Let $M=(\{s\}, \{\lambda, \#, \overline \# \},\{\#, 0\}, \{\mathcal U_V\}, \delta)$  be a $2$-tape CQTM where  $V=\frac{1}{2 \sqrt 2}\left((\ket{\#\#}+\ket{00})\bra{\#\#} \hspace{-0.08cm}+ \hspace{-0.08cm}(\ket{\#\#}-\ket{00})\bra{\#0}\hspace{-0.08cm}+\hspace{-0.08cm}(\ket{\#0}+\ket{0\#})\bra{0\#}\hspace{-0.08cm}+\hspace{-0.08cm}(\ket{\#0}-\ket{0\#})\bra{00}\right),\hfill$   and  $\delta(s,\#)=(h,-,\mathcal U_V)$. If the input is $\ket 0$, then when the machine halts, the state of the cells pointed at by the heads is  entangled: $\frac{1}{\sqrt 2}(\ket{\#0}+\ket{0\#})$. Thus, there is no $1$-tape CQTM which simulates $M$, since entanglement cannot be created by means of one-cell admissible transformations. \end{proof}

\section{CQTM simulates other Quantum Computational Models}
\label{sec:other}

\subsection{Unitary-based quantum computation: Quantum Circuits}

Theorem \ref{thm:cqtm} is a strong evidence of the power and stability of CQTMs: adding a bounded number of tapes to a $2$-tape CQTM does not increase its computational capabilities, and impacts on its efficiency only polynomially. This stability makes $2$-tape CQTMs a good candidate for quantum universality, i.e. the ability to simulate any quantum computation. 
This ability is proved with the following theorem by simulation of any semi-uniform  family of quantum circuits \cite{NO02}. In this section, some basic notions and properties of quantum circuits are given, refer to \cite{Y93,KSV} for fundamentals on quantum circuits.

In the quantum circuit model, the carrier of information is restricted to qubit. Basis states are denoted $\ket 0$ and $\ket 1$. 

\begin{definition}[Quantum Circuits] \label{def:qc} Let $\mathcal G$ be a fixed set of unitary operators (also cal\-led unitary gates.) A $n$-qubit quantum circuit $C_n$ based on  $\mathcal G$ is a sequence $(G_1,R_1), \cdots,$  $(G_s,R_s)$, where $G_i\in \mathcal G$, and $R_i$ is an ordered set of qubits, and $s$ is the size of $C_n$ for $\mathcal G$.
\end{definition}

A uniform family of quantum circuits is a set of circuits $F=\{C_n\}$ such that a classical Turing machine $M_F$ can produce a description of $C_n$ on input $n$ in time $poly(s(n))$, where   $s(n)$ is the size of $C_n$.
 A semi-uniform family of quantum circuits $\{C_n\}$ is a uniform family defined on  a finite set $\mathcal G$ of operators.

\begin{theorem}\label{thm:circuit}
For any semi-uniform family of quantum circuits $F=\{C_n\}$ of size $s$,  there exists a $2$-tape CQTM $M$ operating in time $poly(n,s(n))$, and such that for any $n$-qubit state $\ket{\psi}$, $C_n\ket{\psi}=M(\ket{\psi})$.
\end{theorem}

\begin{proof}
Applying $F=\{C_n\}$ on a $n$-qubit state $\ket{\psi}$ consists in applying the quantum circuit $C_n$ with $\ket{\psi}$ as input. 
Let $\mathcal G$ be a basis of $F$, \emph{i.e.} the set of all the unitary gates used in $F$. 
Since $\mathcal G$ is finite,  $\mathcal G$ has a finite arity $w$, \emph{i.e.} for all $G\in \mathcal G$, the arity of $G$ is less than $w$, where the arity of a gate is the number of qubits on which it operates.

The description of $C_n$ produced by $M_F$ is of the form $(G_1, R_1)\cdots (G_{s(n)},R_{s(n)})$, meaning that $G_i\in \mathcal G$ is applied on the set $\{R_i^{(j)}\}_j$ of qubits in $R_i$, then $G_{i+1}$ is applied, and so on.

Let $M$ be a $(w+1)$-tape CQTM. The admissible transformations of $M$ include the unitary transformation $\mathcal U_{G}$, for all $G\in \mathcal G$. 

A general description of the evolution of $M$ is: 
\begin{itemize}
\item The size $n$ of the input $\ket{\psi}$ located on tape $1$ is computed, using the blank test $\mathcal T_{\#}$ admissible transformation, and stored on tape $T_{w+1}$. This initial stage can be realized within a linear number of steps in $n$.
\item $M_F$ is simulated (see theorem \ref{thm:tm}) and produces a classical  description of  $C_n$ on tape $T_{w+1}$. The complexity of this stage is $p(s(n))$, for some polynomial $p$.
\item For each $(G_i, R_i)$, if $R_i=\{R_i^{(j)}\}_j$, for each $j$, the qubit $R_i^{(j)}$ of tape $1$ is transferred to tape $j$, then $\mathcal U_{G_i}$ is applied, and the qubits of $R_i$ are transferred back to tape $1$. This stage can be realized within $O(ns(n))$ steps.
\end{itemize}

One can show that the resulting state on tape $1$ is the state produced by $C_n\ket{\psi}$. Moreover this simulation can be done within  $ns(n) + p(s(n))$ steps.

Finally, the $(w+1)$-tape CQTM $M$ is simulated by a $2$-tape CQTM $M'$:  the first two stages of $M$, consisting in computing the size of the input and in simulating the classical Turing machine $M_F$ can be simulated without slowdown on $M'$ since only two tapes are required. On the other hand the third stage of $M$ is simulated with a quadratic slowdown  
(see theorem \ref{thm:cqtm}). 

Thus the $2$-tape CQTM $M'$ operates in time $O(n^2s(n)^2+p(s(n)))$. \end{proof}

\begin{corollary}\label{cor:semi}
For any \emph{poly-size} semi-uniform family of quantum circuits $F=\{C_n\}$  there exists a \emph{poly-time} $2$-tape CQTM $M$ such that for any $n$ and any input $\ket{\psi}$ on $n$ qubits, $C_n\ket{\psi}=M(\ket{\psi})$.
\end{corollary}

Semi-uniform families of quantum circuits can be simulated in a polynomial time by means of $2$-tape CQTM. Contrary to semi-uniform families, uniform families of quantum circuits have countable but not necessarily finite basis of gates. 
 Since any CQTM is based on a finite set of admissible transformations, we conjecture that some uniform families of quantum circuits cannot be simulated by means of a CQTM. However, any uniform families of quantum circuits can be approximated:

\begin{theorem}\label{thm:circuitapprox}
For any $\epsilon >0$, and for any uniform family of quantum circuits $F=\{C_n\}$ 
 of size $s$,  
there exists a $2$-tape CQTM $M$ operating in time $poly(n,s(n),1/\epsilon,2^w)$ (where $w$ is the arity of  the gates of $C_n$)  
and such that for any $n$ and any input $\ket{\psi}$ on $n$ qubits, $\vert \vert C_n\ket{\psi}-M(\ket{\psi})\vert \vert  <\epsilon$. \footnote{where $\vert \vert .\vert \vert $ is the Euclidian norm.}
\end{theorem}

\begin{proof}
The proof consists in combining theorem \ref{thm:circuit} and approximation of any unitary transformation using a finite set of unitary transformations.

There exists a finite set of unitary transformations $\mathcal U$ on at most $2$-qubits, such that the approximation within $\epsilon$ of an operation on $w$ qubits takes $poly(1/\epsilon, 2^w)$ gates from $\mathcal U$ \cite{A98}. Moreover there exists an algorithm operating in time $poly(1/\epsilon, 2^w)$ which constructs a description of the circuit realizing the approximation of $U$ \cite{BV97}.

For a given $F=\{C_n\}$ and $\epsilon >0$, let $M$ be a $2$-tape CQTM. A general description of the evolution of $M$ is:
\begin{itemize}
\item The size $n$ of the input $\ket{\psi}$ located on tape $1$ is computed, using the blank test $\mathcal T_{\#}$ admissible transformation, and stored on tape $2$. This initial stage can be realized within a linear number of steps in $n$.
\item $M_F$ is simulated (see theorem \ref{thm:tm}) and produces a classical  description of  $C_n$ on tape $2$. This stage is realized in time $poly(s(n))$.
\item In the description of $C_n$, each gate is replaced by an approximation within $\epsilon/s(n)$ in time $L$, where $L$ is a polynomial in $s(n)/\epsilon$ and $2^w$. This stage is realized in time $Ls(n)$ and the descrition of the circuit is now composed of $poly(s(n),1/\epsilon,2^w)$ gates.
\item For each $(G_i, R_i)$, if $R_i=\{p_i^{(j)}\}_j$, for each $j$, the qubit $R_i^{(j)}$ of tape $1$ is transferred to tape $j$, then $\mathcal U_{G_i}$ is applied, and the qubits of $R_i$ are transferred back to tape $1$. 
\end{itemize}

Thus $M$ operates in time $poly(n,s(n),1/\epsilon,2^w)$. \end{proof}

\begin{corollary}\label{cor:unif}
For any $\epsilon >0$, and for any \emph{poly-size} uniform family of quantum circuits $F=\{C_n\}$ there exists a $2$-tape CQTM $M$ operating in time $poly(n,1/\epsilon, 2^w)$ (where $w$ is the arity of the gates of $C_n$)  and such that for any $n$ and any input $\ket{\psi}$ on $n$ qubits, $\vert \vert C_n\ket{\psi}-M(\ket{\psi})\vert \vert  <\epsilon$. 
\end{corollary}

\subsection{Measurement-based quantum computation : Measurement Calculus}

Alternative models of quantum computation, like the Measurement Calculus \cite{DK04} can also be simulated with $2$-tape CQTMs in  polynomial time.

In the measurement calculus, computations are described by means of \emph{patterns}.
A pattern $\mathcal P$ consists of three finite sets of qubits $V , I, O$, such that $I\subset V$ and $O\subset  V$ and a finite sequence of commands $A_s \cdots A_1$ applying to qubits in $V$.
Qubits in $V\setminus I$ are initialized in the state $\frac{1}{\sqrt 2}(\ket 0 +\ket 1)$, then each command is successively applied. Each command is: 
\begin{itemize}
\item  a two-qubit unitary transformation called controlled-Z, or
\item  a one-qubit measurement, or
\item a  corrective Pauli operator, which is applied or not according to the previous classical outcomes of measurements.
\end{itemize} When all commands in the pattern have been applied,  the output state is the state of the qubits in  $O$.

\begin{lemma} \label{lem:owqc} For any measurement-calculus pattern $\mathcal P$ with commands $A_s \cdots A_1$, acting on $n$ qubits, there exists a $2$-tape CQTM $M_\mathcal P$ acting in time $s^2$, such that for any $\ket \psi$ on $n$ qubits, $\mathcal P(\ket \psi)=M_\mathcal P(\ket \psi)$.
\end{lemma}

\begin{proof}

A general description of the evolution of the $2$-tape CQTM $M_\mathcal P$ simulating $\mathcal P$ is:
\begin{itemize}
\item Qubits corresponding to qubits in $V\setminus I$ are initialized in the state $\frac{1}{\sqrt 2}(\ket 0 +\ket 1)$ within a linear number of steps in $s$.
\item 
Since the sequence of commands is finite, classical internal states of $M_\mathcal P$ are used to describe the sequence of commands. For each command, the head moves first to the addressed qubit on the first tape, in at most $s$ steps, then the command is applied. In the case of controlled-Z, this operation is simulated by transferring one of the two  qubits to the second tape, applying controlled-Z, and transferring back the qubit to the first tape. In this case the number of steps is still linear in $s$. If the command is a one-qubit measurement, the measurement is performed and the classical outcome is stored in the internal state (a bounded memory is enough to store all the classical outcomes since the pattern is finite and each measurement is applied once.) If the command is a corrective Pauli operator, this operator is applied or not, depending on the internal states used to store the previous classical outcomes.
\item Finally, the output qubits corresponding to the set $O$ are reorganized on the first tape in a number of steps  quadratic in $s$.
\end{itemize}
One can show that the resulting state on tape $1$ is $\mathcal P(\ket{\psi})$. This simulation can be done within $s^2$ steps. \end{proof}

In lemma \ref{lem:owqc}, contrary to theorem \ref{thm:circuit} where a semi-uniform family of circuits are simulated, only one pattern, with a fixed input size is simulated. One can extend the model of the measurement calculus, introducing notions of uniform family of patterns, and of semi-uniform family of patterns, similarly to the quantum circuit model. Then,  similar results to those obtained with quantum circuits can be proved.

Since a $2$-tape CQTM naturally simulates unitary-based but also measurement-based models of quantum computation, the classically controlled quantum Turing machine is a unifying model of quantum computation including most of the ingredients which can be used to realize quantum computations: admissible transformations (unitary transformations or projective measurements are nothing but special instances of admissible transformation) ; and classical control  like in the measurement-calculus, which is a model for one-way quantum computation \cite{R03}.

\section{$1$-tape CQTM vs $2$-tape CQTM}
\label{sec:1vs2}

To sum up, two tapes are enough for quantum computation (theorem \ref{thm:circuit}), whereas one tape is enough for classical computation (theorem \ref{thm:tm}) but not for quantum computation (lemma \ref{lem:nuniv}). Thus a \emph{gap} between classical and quantum computations appears. Notice that this result does not contradict the equivalence, in terms of decidability, between classical and quantum computations: the gap appears if and only if quantum data are considered.

One may wonder why $1$-tape CQTMs are not quantum universal whereas Briegel and Raussendorf have proved, with their One-way quantum computer, that one-qubit measurements are universal \cite{R03}. The proof by Briegel and Raussendorf is given with a strong assumption which is that there exists a grid of auxiliary qubits which have been initially prepared, by some unspecified external device, in a globally entangled state (the \emph{cluster state}), whereas the \emph{creation} of entanglement is a crucial point in the proof of lemma \ref{lem:nuniv}. Moreover, another strong assumption of one-way quantum computation is that the input state $\ket{\psi}$ has to be classically known (i.e. a mathematical description of $\ket{\psi}$ is needed), whereas the manipulation of unknown states (i.e. manipulation of qubits in an unknown state) is usual in quantum computation (e.g. teleportation \cite{B93}.) Since none of these assumptions are verified by $1$-tape CQTM, the previous results do not contradict the results of Briegel and Raussendorf.

\section{Measurement-based Quantum Turing Machine}
\label{sec:mqtm}

In the CQTM model the set of admissible transformations $\mathcal A$ represents the quantum resources that are allowed during the computation. By restricting the set $\mathcal A$, one can get specific models of quantum computation like measurement-based quantum computation, or one-way quantum computation. A CQTM $M$ where $\mathcal A$ is restricted to projective measurements only, corresponds to measurement-only quantum computation. It turns out that there is a lack of formal model of quantum computation for alternative models based on measurements. Moreover, classical control plays a crucial role in these promising models of quantum computation.  The restriction of the CQTM to projective measurements produces such a formal model.

\begin{definition}[Measurement-based Quantum Turing Machine] \label{def:mqtm} A Measurement-based quantum Turing machine (MQTM) is a CQTM $M= (K, \Sigma_C, \Sigma_Q, \mathcal A, \delta)$, where $\mathcal A$ is composed of projective measurements only. 
\end{definition}

The following theorem proves that any TM is simulated by a {Las Vegas} MQTM, \emph{i.e.} a MQTM with a non-probabilistic outcome, even if the execution time is unbounded.

\begin{theorem}\label{thm:tmmqtm}
Given any TM $M$ operating in time $f(n)$, where $n$ is the input size, there exists a MQTM $M'$ operating in expected time $O(f(n))$, and such that, for any input $x$, $M(x)=M'(\ket x)$.
\end{theorem}

\begin{proof}
The proof is similar to the simulation of any TM by means of a CQTM (see theorem \ref{thm:tm}), but the permutation $\mathcal P_{[\tau,\sigma]}$ of two symbols, which is a unitary transformation, has to be simulated by means of projective measurements. The simulation consists in applying projective measurement $\mathcal O_{[\tau,\sigma]}$ in the diagonal basis, transforming the state $\ket \tau$ into a uniform superposition $(\ket \tau \pm \ket{\sigma})/\sqrt{2}$. Then the cell is measured according to the standard basis producing $\ket{\sigma}$ with probability $1/2$ and $\ket \tau$ with probability $1/2$. If the simulation of the permutation fails, the sequence of the two measurements $\mathcal O_{[\tau, \sigma]}$ and $Std$ is applied again, untill success. 

Formally, if $M=(K, \Sigma, \delta)$, let $M' \hspace{-0.15cm}=\hspace{-0.15cm}(K', (\Sigma\cup \{\#, \lambda,\top, \bot\}), \Sigma , \{ Std\}\cup \{\mathcal O_{[\tau, \tau']}\}_{ \tau,\tau' \in \Sigma}, \delta')$. For any $(q,\tau) \in K\times \Sigma$, if $\delta(q,\tau)=(p,\sigma,D)$, then $$\begin{array}{rcl}
\delta'(q,\tau)&=&(q_{\tau,\sigma},-,\mathcal O_{[\tau, \sigma]})\\
\delta'(q_{\tau,\sigma},-)&=&(q_\tau',-,Std)\\
\delta'(q'_{\tau,\sigma},\sigma)&=&(p,D,Std)\\
\delta'(q'_{\tau,\sigma}, \tau)&=&(q_{\tau,\sigma},-,\mathcal O_{[\tau,\sigma]})
\end{array}$$

Each transition of $M$ is simulated in expected constant time. \end{proof}

According to theorem \ref{thm:tmmqtm}, unitary transformations are not required to simulate classical computation. One can wonder what is the power of MQTMs compared to CQTMs. In example \ref{ex:h},  a MQTM simulating the application of Hadamard transformation with probability $1$ is described. The simulation is based on state transfer \cite{P04}. More generally, theorem \ref{thm:cqtmmqtm} proves that any CQTM can be efficiently simulated by a MQTM.

\begin{example}[Measurement-based simulation of the Hadamard transformation]\label{ex:h}Consider the problem of simulating a given unitary transformation by means of measurements. We choose to simulate the Hadamard transformation $H$ as an example, on a one-qubit input $\ket{\phi}$, using projective measurements only. Consider the $2$-tape MQTM $M=(K,\Sigma_C,\Sigma_Q,\mathcal{A},\delta)$, with $K=\{s,s',q',q''\}\cup \{ q_{j(a,b)}: a,b \in \{-1,1\}, j=1\ldots 5\}$,
 $\Sigma_Q=\{\#,0,1\}$, 
$\Sigma_C=\{\#, \overline{\#},-1,0,1, \top, \bot\}$,  
$\mathcal{A}=\{\mathcal{T}_{\#},Std,\mathcal{O}_{[\#,0]},\mathcal{O}_{Z},\mathcal{O}_{X},$ $\mathcal{O}_{X\otimes Z},\mathcal{O}_{Z\otimes Z}\}$
 (where $Z$ and $X$ are Pauli matrices)  and $\delta$ is described in figure \ref{fig:h}.
 
 \begin{figure}
 \begin{center}
 $\begin{array}{rccc}
 \hline
 &p\in K,&\tau \in \Sigma_C&\delta(p,\tau)\\
 \hline
 &s & \# & (s',(\rightarrow,-),[-, \mathcal{O}_{[\#,0]}]) \\
 \forall t \in \{-1,0\}  &s'& & (q',(-,-),[-,Std ]) \\
  &q' & \# & (s',(-,-),[-,\mathcal{O}_{O_{[\#,0]}}]) \\
 &q' & 0 & (q'',(-,-),[-, \mathcal{O}_{X}]) \\
\forall i \in \{-1,1\}  &q''& i& (q_{1(i,1)},(-,-),\mathcal{O}_{X\otimes Z} ) \\
\forall i,j \in \{-1,1\}  &q_{1(i,1)}& j& (q_{2(i,j)},(-,-),[\mathcal{O}_{Z}, -] ) \\
\forall i,j,k \in \{-1,1\}  &q_{2(i,j)}& k& (q_{3(i.k,j)},(-,-),[\mathcal{O}_{X}, -] ) \\
\forall k',j,l \in \{-1,1\}  &q_{3(k',j)}& k& (q_{4(k'.l,j)},(-,-),\mathcal{O}_{Z\otimes Z}) \\
\forall l',j,m \in \{-1,1\}  &q_{4(l',j)}& k& (q_{5(l',j.m)},(-,-),[-,\mathcal{O}_{X}]) \\
\forall l',n \in \{-1,1\}: l'=n &q_{5(l',1}& n& (h,(-,-),-) \\
\forall l',m',n \in \{-1,1\}: m'\neq 1 \vee l'\neq n  &q_{5(l',m')}& n& (q_{2(l'.n,m')},(-,-), \mathcal{O}_{Z\otimes Z}) \\

  \hline
 \end{array}$
   \caption{$2$-tape MQTM with projective measurements only for the simulation of $H$}\label{fig:h}
 \end{center}
 \end{figure}
 \begin{figure}
 \begin{center}
 \includegraphics[width=0.8\textwidth]{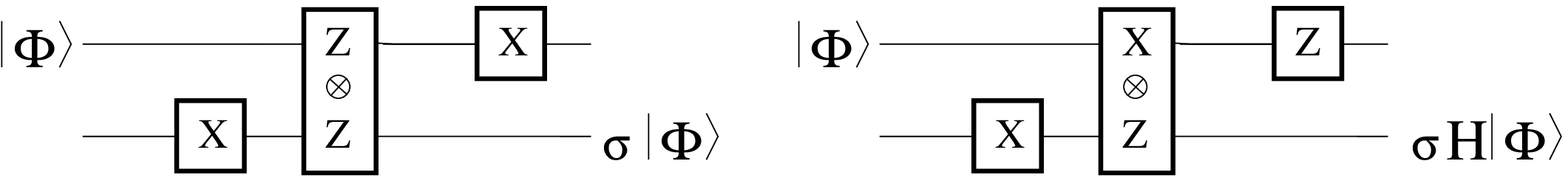} 
 \end{center}
 \caption{\emph{Left:} state transfer - \emph{Right:} generalized state transfer for the simulation of $H$ (see \cite{P04})}\label{fig:st}
 \end{figure}
The input state is placed on tape $T_1$.
The first three transitions are used for transforming the state pointed at by the second head, from $\ket{\#}$ to $\ket{0}$. Then three projective measurements are performed according to generalized state transfer \cite{P04} (see figure \ref{fig:st}): $H\ket{\phi}$, up to a Pauli operator, is placed on $T_2$. Since the result of the computation has to be located on $T_1$, the next three transitions transfer the result of the simulation from $T_2$ to $T_1$. Since state transfer is obtained up to a known Pauli operation (like in teleportation), internal states $q_{j(a,b)}$ are used to memorize the corrective operation: $q_{j(a,b)}$ means that the state transfer is obtained up to the Pauli operator $\sigma_z^{\frac{1-a}{2}}\sigma_x^{\frac{1-b}{2}}$. In order to correct this Pauli operator the state is transferred twice: from $T_1$ to $T_2$, then back from $T_2$ to $T_1$. If $M$ halts then $M(\ket{\phi})=H\ket{\phi}$, but $M$ may never halt, thus $M$ is Las Vegas. Notice that $Pr([M(\ket{\phi})=\nearrow]=0$.
\end{example}

\begin{theorem}\label {thm:cqtmmqtm}
Given any $k$-tape CQTM $M$ operating in time $f(n)$, there exists a $k$-tape MQTM $M'$ operating in time $O(f(n))$, such that for any input $\ket{\psi}$ of size $n$, $M(\ket{\psi})=M'(\ket\psi)$.
\end{theorem}

\begin{proof}
The proof consists of two stages: each admissible transformation used in the transition function of $M$ is transformed into a unitary transformation immediately followed by a projective measurement, then in a second stage, the unitary operation previously obtained is simulated by means of projective measurements.

If $M=(K, \Sigma_C, \Sigma_Q, \mathcal{A}, \delta)$, let $M' =(K', (\Sigma_C^2 \times \{\top, \bot\}), (\Sigma_Q \times \Sigma_C \times \{\top, \bot\}), \mathcal{A'}, \delta')$.

The alphabet of the quantum cells of $M'$ is composed of triplets $(\phi,c,r)$: $\phi \in \Sigma_Q$ is used to simulate the corresponding quantum cell of $M$, $c \in \Sigma_C$ is used in the first stage for transforming  admissible transformations into unitary transformations, and $r \in \{\top, \bot\}$ is an element of the additional workspace needed to simulate any unitary transformations by means of projective measurements. $\mathcal A'$ is a set of admissible transformations acting on quantum states in $\mathcal H_{(\Sigma_Q\times \Sigma_C\times \{\top, \bot\})^k}$. In the following, we  implicitly use that $\mathcal H_{(\Sigma_Q\times \Sigma_C\times \{\top, \bot\})^k}$ is isomorphic to $\mathcal H_{\Sigma_Q^k\times \Sigma_C^k\times \{\top, \bot\}^k}$

\begin{itemize}
\item For  any $(q,c)\in K\times  \Sigma_C$, if $\delta(q,c)=(p,D,A)$, let $\tilde \delta (q,c) = ( q_c, -,\mathcal U_V)$ and $\tilde \delta(q_c,\lambda) = (p, D, \mathcal O_O)$. If $A=\{M_c, c\in \Sigma_C\}$, then $V : \mathcal H_{\Sigma_Q^k \times \Sigma_C^k} \to  \mathcal H_{\Sigma_Q^k \times \Sigma_C^k}$ is a unitary transformation  such that $V \ket{\psi}\ket{\#^k}  = \sum_{c \in \Sigma_C} (M_c\ket \psi) \ket {c \#^{ k-1}} $
 ($V$ can be extended to the case where the second register is not in state $\ket {\#^k}$), and $O$ is a projective measurement of the second register in the basis $\{\ket{c\#^{k-1}}, c \in \Sigma_C\}$. If $\delta$ is replaced by $\tilde \delta$ an exact simulation of $M$ is obtained.

\item Let $R = V \otimes \ket {\bot} \bra {\top}\otimes Id_{\mathcal H_{\{\top, \bot\}^{k-1}}} + V^\dagger \otimes \ket {\top} \bra {\bot}\otimes Id_{\mathcal H_{\{\top, \bot\}^{k-1}}}$, $P_\top = (Id_{\mathcal H_{\Sigma_Q^k\times \Sigma_C^k\times \{\top, \bot\}^k}}+R)/2$, and $P_\bot=(Id_{\mathcal H_{\Sigma_Q^k\times \Sigma_C^k\times \{\top, \bot\}^k}}-R)/2$. $\mathcal V= \{P_\top, P_\bot\}$ is a projective measurement.  
One can show that for any $\ket \phi \in  \mathcal H_{\Sigma_Q^k \times \Sigma_C^k}$, a measurement of $\ket \phi \otimes \ket {\top^k}$ according to $\mathcal V$, followed by a projective measurement according to $L=\{Id_{\mathcal H_{\Sigma_Q^k \times \Sigma_C^k}}\otimes \ket{\top} \bra {\top}\otimes Id_{\mathcal H_{\{\top, \bot\}^{k-1}}}, Id_{\mathcal H_{\Sigma_Q^k \times \Sigma_C^k}}\otimes \ket \bot \bra \bot\otimes Id_{\mathcal H_{\{\top, \bot\}^{k-1}}}\}$, produces $(V\ket \phi)\otimes \ket {\bot \top^{k-1}}$ with probability $1/2$, and $\ket \phi \otimes \ket {\top^k}$ with probability $1/2$. Then, for any  $(q,c)\in K\times  \Sigma_C$, if $\tilde \delta (q,c) = ( q_c, -,\mathcal U_V)$ and $\tilde \delta(q_c,\lambda) = (p, D, \mathcal O_O)$, let $\delta'$ be such that:
 $$\begin{array}{rcl}
 \delta'(q,c)&=&(q_c',-, \mathcal O_{\mathcal V})\\
  \delta'(q_c',-)&=&(q_c'',-, \mathcal O_{L})\\
  \delta'(q_c'', \top)&=&(q_c',-, \mathcal O_{\mathcal V})\\
   \delta'(q_c'', \bot)&=&(p,D, \mathcal O_{O})\\
 \end{array}$$ 
\end{itemize}

Thus, any transition of the CQTM $M$ can be simulated on the MQTM $M'$ within an expected constant number of transitions. As a consequence, $M'$ simulates $M$ with a expected execution time $O(f(n))$. \end{proof}

Since any $k$-CQTM is efficiently simulated by a $k$-tape MQTM, the results proved for CQTMs in previous sections hold for MQTMs also.

\section{Conclusion}
This paper introduces classically-controlled quantum Turing machines (CQTM), a new abstract model for quantum computations. 
This model allows a rigorous formalization of the inherent interactions between the quantum world and the classical world during a quantum computation. Any classical Turing machine is simulated by a CQTM without loss of efficiency, moreover any $k$-tape CQTM is simulated by a $2$-tape CQTM affecting the execution time only polynomially. 

Moreover a gap between classical and quantum computations is pointed out in clas\-sically-controlled quantum computations.

MQTMs, a restriction of CQTMs to projective measurements, are introduced. Any CQTM can be efficiently simulated by a MQTM. This formally proves the power of alternative models based on measurements only \cite{N01,L03,P04}. 

The classically-controlled quantum Turing machine is a good candidate for establishing a bridge between, on one side, theoretical models like QTM and MQTM, and on the other side practical models of quantum computation like quantum random access machines.

\end{document}